\documentclass[preprint,showpacs,showkeys,preprintnumbers,amsmath,amssymb,aps]{revtex4-1}

\usepackage{bm,amssymb,amsmath,mathrsfs}
\usepackage[usenames]{color}

\usepackage{dcolumn}
\usepackage[pdftex]{graphicx}

\newcommand{\bc}{\begin{center}}
\newcommand{\ec}{\end{center}}
\newcommand{\bit}{\begin{itemize}}
\newcommand{\eit}{\end{itemize}}
\newcommand{\bq}{\begin{equation}}
\newcommand{\eq}{\end{equation}}

\begin{document}

\title{Orbit and spin tracking in a radial electric field}

\author{S.~R.~Mane}
\email{srmane001@gmail.com}

\affiliation{Convergent Computing Inc., P.~O.~Box 561, Shoreham, NY 11786, USA}

\begin{abstract}
I derive an expression for the spin decoherence rate for a smooth focusing electrostatic ring with a radial electric field.
The answer contains some salutary lessons about spin decoherence calculations in electrostatic storage rings:
the numerical tracking results disagreed with na{\"\i}ve analytical calculations,
but it was the tracking results which were correct, and more careful analytical calculations (canonical transformations) were required.
\end{abstract}


\keywords{electric dipole moment, storage ring, spin coherence, canonical transformations
\vskip 0.5in
{\em
A shorter version of this material was presented at an EDM workshop in Trento in October 2012.
I realize by now that I do not have the motivation to write this up formally for a paper in a peer reviewed journal,
and it is simply wearisome to explain undergraduate concepts of potential and kinetic energy to the EDM collaboration {\em again}.
There is nothing in this document beyond careful and patient application of canonical transformations
of Hamiltonian dynamics, which can be found in any graduate textbook on higher analytical dynamics.
}
}

\maketitle

\section{Introduction}
I tracked the orbit and spin in a model of a homogenous radial electric field $E \propto 1/r$.
I treated only motion in the horizontal plane, and only for on-energy particles $H=H_0 = \gamma_0 mc^2$.
Hence there was no dispersion orbit:
the orbital motion consisted purely of betatron oscillations around the reference radius $r_0$.
The problem was more subtle than I anticipated.
My tracking output did {\em not} agree with averages I had previously calculated for various parameters.
Significantly, the tracking output indicated that $\langle x_\beta\rangle \ne 0$ for this model.
I realized eventually that this is correct:
my tracking program is reliable and was sending me a message.
I calculated the value of $\langle x_\beta/r_0\rangle$ to $O(x_0^2/r_0^2)$
using canonical transformations (see below).
I then recalculated the averages for various parameters,
and I then obtained excellent agreement with the tracking output.

\vskip 0.1in
In what follows I shall define the parameter $\alpha$, 
the angle between the spin and the direction of the velocity 
(basically $\cos\alpha$ is the helicity), 
and the na{\"\i}ve analytical expression for the average $\langle d\alpha/d\theta\rangle$.
The tracking output did not agree with this expression.
I shall then derive a more detailed expression for $\langle x_\beta/r_0\rangle$,
and then rederive a more accurate expression for $\langle d\alpha/d\theta\rangle$.

\vskip 0.1in
The fact that $\langle x_\beta/r_0\rangle \ne 0$ for this model is food for thought 
about assumptions which have been made in various statistical analyses of EDM averages.

\section{Helicity}
Use cylindrical polar coordinates $(\rho,\theta,z)$, where the arc-length along the reference orbit is $s=r_0\theta$. 
The electric field is radial and the field and potential are given by
\bq
\bm{E} = -E_0\,\frac{r_0}{\rho}\,\hat{\bm{\rho}} \,,\qquad
V(r) = E_0 r_0\,\ln\frac{\rho}{r_0} \,.
\eq
First of all we know that for this model, all circular orbits circulate with
the same kinetic energy and different potential energies. 
We have the relation
\bq
eE_0r_0 = p_0v_0 = mc^2 \gamma_0\beta_0^2 \,.
\eq
If $\bm{B}=0$ (no magnetic field), 
the spin precession equation of motion for the helicity is given by
\bq
\frac{d(\bm{s}\cdot\hat{\bm{\beta}})}{dt} = \frac{e}{mc}\biggl(a - \frac{1}{\beta^2\gamma^2}\biggr)
(\bm{\beta}\times\bm{E})\cdot(\bm{s}\times\hat{\bm{\beta}})\,.
\eq
We neglect vertical motion. The orbit and spin precess in the horizontal plane.
We set 
\bq
\bm{s}\cdot\hat{\bm{\beta}} = \cos\alpha \,,\qquad
\bm{s}\times\hat{\bm{\beta}} = \sin\alpha\,\hat{\bm{z}} \,.
\eq
Then, using $c\beta_\theta = v_\theta = \rho\dot{\theta}$,
\bq
\begin{split}
\frac{d\alpha}{dt} &= -\frac{e}{mc}\biggl(a - \frac{1}{\beta^2\gamma^2}\biggr) (\bm{\beta}\times\bm{E})\cdot\hat{\bm{z}}
\\
&= \frac{e}{mc^2}\biggl(a - \frac{1}{\beta^2\gamma^2}\biggr) \rho\dot{\theta}\,\frac{E_0r_0}{\rho} (\hat{\bm{\theta}}\times\hat{\bm{\rho}})\cdot\hat{\bm{z}}
\\
&= -\frac{eE_0r_0}{mc^2}\biggl(a - \frac{1}{\beta^2\gamma^2}\biggr) \dot{\theta}\,.
\end{split}
\eq
Hence
\bq
\frac{d\alpha}{d\theta} = -\gamma_0\beta_0^2 \biggl(a - \frac{1}{\beta^2\gamma^2}\biggr) \,.
\eq
This is quite general. 
It is applicable for any value of the reference momentum.
However, the case of interest to us is when the reference is at the magic momentum, given by $a = 1/(\beta_0^2\gamma_0^2)$.
Then we must expand $\beta\gamma$ to include the off-axis motion.

\vskip 0.1in
{\em Why is $\alpha$ relevant?} 
I have my own personal opinion about this, but I shall postpone that to a later document.

\section{Na{\"\i}ve average}
The quantity of interest is the secular average $\langle d\alpha/d\theta\rangle$.
For brevity define $\xi=x/r_0$. Then using $H = \gamma mc^2 + \Phi$,
\bq
\gamma = \frac{H}{mc^2} - \frac{\Phi}{mc^2} \,.
\eq
In this note I treat only the case $H = H_0 = \gamma_0 mc^2$.
Note also that 
\bq
\Phi = eE_0r_0\ln\Bigl(1+\frac{x}{r_0}\Bigr) = mc^2\gamma_0\beta_0^2 \ln(1+\xi)
\simeq mc^2\gamma_0\beta_0^2 \Bigl(\xi - \frac{\xi^2}{2}\Bigr) \,.
\eq
Then
\bq
\label{eq:gamma}
\gamma = \gamma_0\,\bigl[\, 1 - \beta_0^2\ln(1+\xi)\,\bigr]  \,.
\eq
Then, using $\beta^2\gamma^2 = \gamma^2-1$,
\bq
\begin{split}
1 - \frac{\beta_0^2\gamma_0^2}{\beta^2\gamma^2} =
1 - \frac{\beta_0^2\gamma_0^2}{\gamma^2 -1}
&= 1 - \frac{\beta_0^2\gamma_0^2}{\gamma_0^2(1-\beta_0^2\ln(1+\xi))^2 -1}
\\
&= 1 - \frac{\beta_0^2\gamma_0^2}{\beta_0^2\gamma_0^2(1-2\ln(1+\xi)+\beta_0^2\ln^2(1+\xi))}
\\
&= 1 - \frac{1}{1-2\ln(1+\xi)+\beta_0^2\ln^2(1+\xi)}
\\
&\simeq 1 - \frac{1}{1-2\xi +(1+\beta_0^2)\xi^2}
\\
&\simeq -2\xi +(1+\beta_0^2)\xi^2 - 4\xi^2 
\\
&= -2\xi -(3-\beta_0^2)\xi^2 \,.
\end{split}
\eq
Then, setting $a = 1/(\beta_0^2\gamma_0^2)$,
\bq
\frac{d\alpha}{d\theta} = -\gamma_0\beta_0^2 \biggl(a - \frac{1}{\beta^2\gamma^2}\biggr) 
= -\frac{1}{\gamma_0} \biggl(1 - \frac{\beta_0^2\gamma_0^2}{\beta^2\gamma^2}\biggr)
\simeq \frac{1}{\gamma_0} \biggl[\, 2\xi +(3-\beta_0^2)\xi^2 \,\biggr]\,.
\eq
Then, assuming $\langle \xi\rangle = 0$ and $\langle \xi^2\rangle = (1/2)(x_0^2/r_0^2)$, we obtain
\bq
\label{eq:dadtnaive}
\biggl\langle\frac{d\alpha}{d\theta}\biggr\rangle 
\simeq \frac{3-\beta_0^2}{2\gamma_0} \,\frac{x_0^2}{r_0^2} \,.
\eq
The tracking results disagreed with this significantly.
The above result indicates that for $\beta_0\to0$ (so $\gamma_0\to1$),
then $\langle d\alpha/d\theta\rangle \to (3/2)(x_0^2/r_0^2)$.
However the tracking results indicated that 
$\langle d\alpha/d\theta\rangle \propto \beta_0^4$ instead.

\vskip 0.1in
The tracking results also indicated that $\langle \xi\rangle = O(x_0^2/r_0^2) \ne 0$.
I eventually realized that this is correct, and that it is the key.

\section{Hamiltonian \&\ canonical transformations}
Henceforth I set $c=1$.
The Hamiltonian in cylindrical coordinates $(\rho,\theta,z)$ is 
\bq
\label{eq:hamtim}
H = \biggl[\, m^2 + p_\rho^2 + \frac{p_s^2}{(\rho/\rho_0)^2} +p_z^2 \,\biggr]^{\frac12} + \Phi \,.
\eq
We set $p_z=0$ below and treat only motion in the horizontal plane.
Define $p_0=m\gamma_0\beta_0$. The potential is
\bq
\Phi = eE_0r_0\ln\frac{\rho}{\rho_0} = m\gamma_0\beta_0^2\ln\frac{\rho}{\rho_0} \,.
\eq
Here $\rho=r_0+x$ and $\rho_0=r_0$.
The logarithmic potential means that all of the orbits are bounded.
The equation for the classical turning radius $x_*$ is easy to write down.
We set $\rho=\rho_*$ and $p_\rho=0$ in eq.~\eqref{eq:hamtim}. 
Then, for some $H=H_*$ and $p_s=L/\rho_0$ ($L$ is the angular momentum), the classical turning radius is the solution of the equation
\bq
H_* = \biggl[\, m^2 + \frac{L^2}{\rho_*^2} \,\biggr]^{\frac12} + eE_0r_0\ln\frac{\rho_*}{\rho_0} \,.
\eq
In general there are two solutions for $\rho_*$: there is both a maximum and a minimum classical turning radius.
If the maximum and minimum classical turning radii are equal then the orbit is a circle.
However, when using a Taylor series and perturbation theory, 
I shall expand to the fourth order in $\xi=x/r_0$. Then
\bq
\Phi = eE_0r_0\ln(1+\xi) \simeq eE_0r_0\,\biggl[\, \xi-\frac{\xi^2}{2}+\frac{\xi^3}{3}-\frac{\xi^4}{4} \,\biggr] \,.
\eq
The highest power is $-\frac14\xi^4$, which is {\em negative},
hence orbits can escape to infinity.
This simply means the perturbation series is then not valid, and the higher order terms are not negligible.
There is no infinity in the actual orbital motion.

We use the arc-length $s$ along the reference orbit as the independent variable.
Then
\bq
\label{eq:kcantran}
K = -p_s = -\frac{\rho}{\rho_0}\,\biggl[\, (H-\Phi)^2 - m^2 - p_x^2 \,\biggr]^{\frac12} \,.
\eq
Let us employ canonical transformations and diagonalize the Hamiltonian, by expanding terms in a Taylor series.
Note that $x$ always appears in the dimensionless combination $x/r_0$, hence use $\xi=x/r_0$. 
To preserve the Hamiltonian structure of the equations, 
we scale the independent variable to $\theta=s/r_0$:
\bq
\frac{d\xi}{d\theta} = \frac{dx}{ds} = \frac{\partial K}{\partial p_x} \,,\qquad
\frac{dp_x}{d\theta} = r_0\,\frac{dp_x}{ds} = -\frac{\partial K}{\partial \xi} \eq
Next, we scale the momentum $p_x = p_0\,p_\xi$.
To preserve the Hamiltonian structure of the equations, we divide $K$ by $p_0$:
\bq
\frac{d\xi}{d\theta} = \frac{\partial (K/p_0)}{\partial p_\xi} \,,\qquad
\frac{dp_\xi}{d\theta} = -\frac{\partial (K/p_0)}{\partial \xi} \,.
\eq
Hence we define $\bar{K}=K/p_0$.
I shall treat only the case $H = \gamma_0 m$ below.
There is no off-energy dispersion orbit.
Then $x=p_x=0$ is a solution, viz.~motion in a circle of radius $r_0$.
The orbital motion consists purely of betatron oscillations.
Then
\bq
\begin{split}
\bar{K} = \frac{K}{p_0} = -\frac{p_s}{p_0} &= 
-\frac{1+\xi}{p_0}\,\biggl[\, m^2\gamma_0^2\bigl[\,1 -\beta_0^2\ln(1+\xi)\,\bigr]^2 - m^2 - p_0^2p_\xi^2 \,\biggr]^{\frac12} 
\\
&= -(1+\xi)\,\biggl[\, 1 -2\ln(1+\xi) + \beta_0^2\ln^2(1+\xi) - p_\xi^2 \,\biggr]^{\frac12} \,.
\end{split}
\eq
Define $\kappa = p_s/p_0$.
Note that $K$ is invariant along an orbit, hence $p_s$ and hence $\kappa$ are also invariant. Hence $p_s=p_{s0}$ and $\kappa=\kappa_0$.
Note that the value of $\kappa$ must be precomputed using the initial data.
We can employ the equivalent Hamiltonian
\bq
K_1 = \frac{1}{2\kappa}(1+\xi)^2\,\biggl[\, p_\xi^2 -1 +2\ln(1+\xi) -\beta_0^2\ln^2(1+\xi) \,\biggr] \,.
\eq
The partial derivatives of all dynamical variables are the same using $K_1$ and $\bar{K}$.
We work with $K_1$ below. 
We expand $K_1$ in a Taylor series in powers of $\xi$, up to the fourth power:
\bq
\begin{split}
K_1 &\simeq \frac{1}{2\kappa}(1+2\xi+\xi^2) \,
\biggl[\, p_\xi^2 - 1 +2\xi -\xi^2 +\frac23\,\xi^3 -\frac12\,\xi^4
-\beta_0^2\Bigl(\xi^2 -\xi^3 +\frac{11}{12}\xi^4\Bigr)\,\biggr]
\\
&\simeq \frac{1}{2\kappa} \,
\biggl[\, p_\xi^2(1+2\xi+\xi^2) - 1 +(2-\beta_0^2)\xi^2 
+(\frac23-\beta_0^2)\xi^3 -\frac{1}{12}(2-\beta_0^2)\xi^4
\,\biggr] 
\\
&= \textrm{(const.)} 
+\frac{p_\xi^2}{2\kappa} + \frac{\kappa}{2}\,\frac{2-\beta_0^2}{\kappa^2}\,\xi^2
+\frac{1}{2\kappa} \,
\biggl[\, p_\xi^2(2\xi+\xi^2) +(\frac23-\beta_0^2)\xi^3 -\frac{1}{12}(2-\beta_0^2)\xi^4 \,\biggr] \,.
\end{split}
\eq
We discard the constant term, and separate $K_1$ into quadratic and anharmonic terms. 
The quadratic terms describe the motion of a particle of mass $\kappa$,
with a tune $\nu_x = \sqrt{2-\beta_0^2}/\kappa$. 
We can define action-angle variables $(J,\phi)$ for the linear dynamical motion
\bq
\label{eq:aalin}
\xi = \sqrt{2J/(\kappa\nu_x)}\,\cos\phi \,,\qquad
p_\xi = -\sqrt{2J\kappa\nu_x}\,\sin\phi \,,\qquad 
\frac{d\phi}{d\theta} = \nu_x \,.
\eq
Then the Hamiltonian in (linear dynamical) action-angle variables is
\bq
\begin{split}
\label{eq:kaa}
\mathcal{K} = \nu_x J 
&+\nu_x J\sin^2\phi\,\biggl[\,2\Bigl(\frac{2J}{\kappa\nu_x}\Bigr)^{1/2}\,\cos\phi +\frac{2J}{\kappa\nu_x}\,\cos^2\phi\,\biggl]
\\
&+\frac{1}{2\kappa}(\frac23-\beta_0^2)
\Bigl(\frac{2J}{\kappa\nu_x}\Bigr)^{3/2}\,\cos^3\phi
-\frac{J^2}{6\kappa}\,\cos^4\phi \,.
\end{split} 
\eq
We use the trigonometric identities
\begin{subequations}
\begin{align}
\sin^2\phi\cos\phi &= \frac12\sin(2\phi)\sin\phi 
= \frac14\,\bigl[\, \cos\phi - \cos(3\phi) \,\bigr] \,,
\\
\sin^2\phi\cos^2\phi &= \frac14\sin^2(2\phi)
= \frac18\,\bigl[\, 1 - \cos(4\phi) \,\bigr] \,,
\\
\cos^3\phi &= \frac14\,\bigl[\,3\cos\phi +\cos(3\phi)\,\bigr] \,,
\\
\cos^4\phi &= \frac18\,\bigl[\,3+4\cos(2\phi)+\cos(4\phi) \,\bigr] \,.
\end{align}
\end{subequations}
Then the expansion in Fourier harmonics is
\bq
\begin{split}
\label{eq:kaa1}
\mathcal{K}
= \nu_x J 
&
+\frac{\kappa\nu_x^2}{4}\,\Bigl(\frac{2J}{\kappa\nu_x}\Bigr)^{3/2}\,\bigl[\, \cos\phi - \cos(3\phi) \,\bigr]
\\
& +\frac{J^2}{4\kappa}\,\bigl[\, 1 - \cos(4\phi) \,\bigr]
\\
& +\frac{3\kappa^2\nu_x^2-4}{24\kappa}
\Bigl(\frac{2J}{\kappa\nu_x}\Bigr)^{3/2}\,\bigl[\,3\cos\phi +\cos(3\phi) \,\bigr]
\\
& -\frac{J^2}{48\kappa}\,\bigl[\, 3 +4\cos(2\phi) +\cos(4\phi)\,\bigr] 
\\
= \nu_x J 
& +\Bigl(\frac{2J}{\kappa\nu_x}\Bigr)^{3/2}\,\biggl\{ \frac{\kappa\nu_x^2}{4}\,\bigl[\, \cos\phi - \cos(3\phi) \,\bigr]
+\frac{3\kappa^2\nu_x^2-4}{24\kappa}\,\bigl[\,3\cos\phi +\cos(3\phi) \,\bigr] \biggr\}
\\
& +\frac{J^2}{48\kappa}\,\bigl[\, 9 -4\cos(2\phi) - 13\cos(4\phi) \,\bigr] \,.
\end{split}
\eq
We eliminate the terms in $J^{3/2}$, all of which are nonsecular.
Let the new action-angle variables be $(J_1,\phi_1)$.
The generating function is
\bq
\label{eq:gf1}
\mathcal{G}_1 = \phi J_1 
-\Bigl(\frac{2J_1}{\kappa\nu_x}\Bigr)^{3/2}\,\biggl\{ \frac{\kappa\nu_x}{4}\,\bigl[\, \sin\phi - \frac13\,\sin(3\phi) \,\bigr]
+\frac{3\kappa^2\nu_x^2-4}{24\kappa\nu_x}\,\bigl[\,3\sin\phi +\frac13\,\sin(3\phi) \,\bigr] \biggr\} \,.
\eq
The new angle variable $\phi_1$ is given by
\bq
\begin{split}
\phi_1 &= \frac{\partial\mathcal{G}_1}{\partial J_1}
\\
&= \phi 
-\Bigl(\frac{2}{\kappa\nu_x}\Bigr)^{3/2}\,J_1^{1/2}\,\biggl\{\frac{3\kappa\nu_x}{8}\,\bigl[\, \sin\phi - \frac13\,\sin(3\phi) \,\bigr]
+\frac{3\kappa^2\nu_x^2-4}{16\kappa\nu_x} \,\bigl[\,3\sin\phi +\frac13\,\sin(3\phi) \,\bigr] \biggr\} \,.
\end{split}
\eq
The old action variable $J$ is given by
\bq
J = \frac{\partial\mathcal{G}_1}{\partial\phi}
= J_1 
-\Bigl(\frac{2J_1}{\kappa\nu_x}\Bigr)^{3/2}\, \biggl\{ \frac{\kappa\nu_x}{4}\,\bigl[\, \cos\phi - \cos(3\phi) \,\bigr]
+\frac{3\kappa^2\nu_x^2-4}{24\kappa\nu_x} \,\bigl[\,3\cos\phi +\cos(3\phi) \,\bigr] \biggr\} \,.
\eq
Then we may set $J \simeq J_1$ in the $O(J^2)$ terms in $\mathcal{K}$.
For the $O(J^{3/2})$ terms in $\mathcal{K}$, we may set $\phi \simeq \phi_1$, which yields
\bq
\begin{split}
J^{3/2} &= J_1^{3/2} \,\Bigl( 1 + \frac{\Delta J_1}{J_1}\Bigr)^{3/2} 
\\
&\simeq  J_1^{3/2} + \frac32 J_1^{1/2} \Delta J_1
\\
&\simeq J_1^{3/2}  
-\frac32\Bigl(\frac{2}{\kappa\nu_x}\Bigr)^{3/2}\,J_1^2\,\biggl\{ \frac{\kappa\nu_x}{4}\,\bigl[\, \cos\phi_1 - \cos(3\phi_1) \,\bigr]
+\frac{3\kappa^2\nu_x^2-4}{24\kappa\nu_x} \,\bigl[\,3\cos\phi_1 +\cos(3\phi_1) \,\bigr] \biggr\} \,.
\end{split}
\eq
The transformed Hamiltonian is (note that $\nu_x\Delta J_1$ cancels the terms in $J_1^{3/2}$)
\bq
\begin{split}
\mathcal{K}_1 &= \mathcal{K} +\underbrace{\frac{\partial\mathcal{G}_1}{\partial\theta}}_{=0}
\\
&\simeq \nu_x (J_1+\Delta J_1) 
+\frac{J_1^2}{48\kappa}\,\bigl[\, 9 -4\cos(2\phi_1) - 13\cos(4\phi_1) \,\bigr]
\\
& \quad
+\biggl\{ \frac{\kappa\nu_x^2}{4}\,\bigl[\, \cos\phi - \cos(3\phi) \,\bigr]
+\frac{3\kappa^2\nu_x^2-4}{24\kappa} \,\bigl[\,3\cos\phi +\cos(3\phi) \,\bigr] \biggr\}
\,\Bigl(\frac{2}{\kappa\nu_x}\Bigr)^{\frac32}\,J_1^{\frac32}\Bigl(1+\frac{\Delta J_1}{J_1}\Bigr)^{\frac32}
\\
&
\\
&\simeq \nu_x J_1 +\frac{3}{16\kappa}J_1^2
-\frac{J_1^2}{48\kappa}\,\bigl[\, 4\cos(2\phi_1) + 13\cos(4\phi_1) \,\bigr]
\\
& \qquad\quad
+\frac{3}{2\kappa}\,\biggl\{\frac{5\kappa^2\nu_x^2-4}{8}\, \cos\phi_1 -\frac{3\kappa^2\nu_x^2+4}{24}\, \cos(3\phi_1) \biggr\} 
\,\Bigl(\frac{2}{\kappa\nu_x}\Bigr)^{\frac32}\,J_1^{\frac12}\Delta J_1 
\\
&
\\
&\simeq \nu_x J_1 +\frac{3}{16\kappa}J_1^2
-\frac{J_1^2}{48\kappa}\,\bigl[\, 4\cos(2\phi_1) + 13\cos(4\phi_1) \,\bigr]
\\
& \qquad\qquad
-\frac{J_1^2}{48\kappa^5\nu_x^4}\,
\biggl\{3(5\kappa^2\nu_x^2-4)\, \cos\phi_1 -(3\kappa^2\nu_x^2+4)\, \cos(3\phi_1) \biggr\}^2 
\\
\\
&\simeq \nu_x J_1 +\frac{J_1^2}{96\kappa^5\nu_x^4} \, \biggl[\, 18\kappa^4\nu_x^4
- 9(5\kappa^2\nu_x^2-4)^2 - (3\kappa^2\nu_x^2+4)^2 \,\biggr]
+\textrm{(oscillatory)} \,.
\end{split}
\eq
From this we can deduce the leading order tuneshift
\bq
\nu_{x1} = \frac{\partial\mathcal{K}_1}{\partial J_1} 
= \nu_x +\frac{J_1}{48\kappa^5\nu_x^4} \, \biggl[\, 18\kappa^4\nu_x^4
- 9(5\kappa^2\nu_x^2-4)^2 - (3\kappa^2\nu_x^2+4)^2 \,\biggr] \,.
\eq
Our real interest is in the value of $\langle x\rangle$. Now 
\bq
\begin{split}
\xi &= \Bigl(\frac{2J}{\kappa\nu_x}\Bigr)^{1/2}\,\cos\phi
\\
&= \Bigl(\frac{2}{\kappa\nu_x}\Bigr)^{1/2}J_1^{1/2}\Bigl(1+\frac{\Delta J_1}{J_1}\Bigr)^{1/2}\,\cos(\phi_1+\Delta\phi_1)
\\
&\simeq \Bigl(\frac{2}{\kappa\nu_x}\Bigr)^{1/2}\bigl(J_1^{1/2}+\frac12J_1^{-1/2}\Delta J_1\bigr)\,
\Bigl[\,\cos\phi_1\cos(\Delta\phi_1) - \sin\phi_1\sin(\Delta\phi_1)\,\Bigr]
\\
&\simeq \Bigl(\frac{2}{\kappa\nu_x}\Bigr)^{1/2}\bigl(J_1^{1/2}+\frac12J_1^{-1/2}\Delta J_1\bigr)\,
\Bigl[\,\cos\phi_1 - \Delta\phi_1\sin\phi_1\,\Bigr]
\\
&\simeq \Bigl(\frac{2J_1}{\kappa\nu_x}\Bigr)^{1/2} \, \cos\phi_1
\\
&\qquad
-\frac{2J_1}{\kappa^2\nu_x^2}\, \biggl[\, \frac{\kappa\nu_x}{4}\,\bigl[\, \cos\phi_1 - \cos(3\phi_1) \,\bigr]
+\frac{3\kappa^2\nu_x^2-4}{24\kappa\nu_x} \,\bigl[\,3\cos\phi_1 +\cos(3\phi_1) \,\bigr] \biggr]\, \cos\phi_1
\\
&\qquad
-\frac{6J_1}{\kappa^2\nu_x^2}\,\biggl[\,\frac{\kappa\nu_x}{4}\,\bigl[\, \sin\phi_1 - \frac13\,\sin(3\phi_1) \,\bigr]
+\frac{3\kappa^2\nu_x^2-4}{24\kappa\nu_x} \,\bigl[\,3\sin\phi_1 +\frac13\,\sin(3\phi_1) \,\bigr] \biggr]\, \sin\phi_1 \,.
\end{split}
\eq
We want the average $\langle \xi\rangle$, which is given by the secular terms 
\bq
\langle \xi\rangle = -\frac{5\kappa^2\nu_x^2-4}{2\kappa^3\nu_x^3}\,J_1 \,.
\eq
This expression is valid to $O(x_0^2/r_0^2)$.
To this level of approximation, the initial conditions yield  
\bq
J_1 \simeq \frac{\kappa\nu_x}{2}\, \frac{x_0^2}{r_0^2} \,.
\eq
Then
\bq
\label{eq:xiavg}
\langle \xi\rangle = -\frac{5\kappa^2\nu_x^2-4}{4\kappa^2\nu_x^2}\,\frac{x_0^2}{r_0^2}
= -\frac14\,\frac{6-5\beta_0^2}{2-\beta_0^2}\,\frac{x_0^2}{r_0^2} \,.
\eq
Let us also calculate $\langle p_\xi\rangle$ as a sanity check. 
Because the Hamiltonian is invariant under a change of sign $p_\xi \to -p_\xi$,
we must have $\langle p_\xi\rangle = 0$.
We obtain
\bq
\begin{split}
p_\xi &= -\sqrt{2J\kappa\nu_x}\,\sin\phi
\\
&= -\sqrt{2\kappa\nu_x}\,J_1^{1/2}\Bigl(1+\frac{\Delta J_1}{J_1}\Bigr)^{1/2}\,\sin(\phi_1+\Delta\phi_1)
\\
&\simeq -\sqrt{2\kappa\nu_x}\,\bigl(J_1^{1/2}+\frac12J_1^{-1/2}\Delta J_1\bigr)\,
\Bigl[\,\sin\phi_1\cos(\Delta\phi_1) + \cos\phi_1\sin(\Delta\phi_1)\,\Bigr]
\\
&\simeq -\sqrt{2\kappa\nu_x}\,\bigl(J_1^{1/2}+\frac12J_1^{-1/2}\Delta J_1\bigr)\,\Bigl[\,\sin\phi_1 + \Delta\phi_1\cos\phi_1\,\Bigr]
\\
&\simeq -\sqrt{2J_1\kappa\nu_x} \, \sin\phi_1
\\
&\qquad
+\frac{2J_1}{\kappa\nu_x}\, \biggl[\, \frac{\kappa\nu_x}{4}\,\bigl[\, \cos\phi_1 - \cos(3\phi_1) \,\bigr]
+\frac{3\kappa^2\nu_x^2-4}{24\kappa\nu_x} \,\bigl[\,3\cos\phi_1 +\cos(3\phi_1) \,\bigr] \biggr]\, \sin\phi_1
\\
&\qquad
-\frac{6J_1}{\kappa\nu_x}\,\biggl[\,\frac{\kappa\nu_x}{4}\,\bigl[\, \sin\phi_1 - \frac13\,\sin(3\phi_1) \,\bigr]
+\frac{3\kappa^2\nu_x^2-4}{24\kappa\nu_x}
\,\bigl[\,3\sin\phi_1 +\frac13\,\sin(3\phi_1) \,\bigr] \biggr]\, \cos\phi_1 \,.
\end{split}
\eq
There are no secular terms so $\langle p_\xi\rangle=0$ as required.

\section{Helicity at magic momentum}
We know that
\bq
\frac{d\alpha}{d\theta} = -\gamma_0\beta_0^2 \biggl(a - \frac{1}{\beta^2\gamma^2}\biggr) 
= -\frac{1}{\gamma_0} \biggl(1 - \frac{\beta_0^2\gamma_0^2}{\gamma^2-1}\biggr) \,.
\eq
We also know that
\bq
\frac{d\alpha}{d\theta} \simeq \frac{1}{\gamma_0} \bigl[\, 2\xi +(3-\beta_0^2)\xi^2 \,\bigr] 
\eq
We now know that to $O(x_0^2/r_0^2)$
\bq
\xi \simeq \frac{x_0}{r_0}\,\cos(\nu_x\phi) 
-\frac14\,\frac{6-5\beta_0^2}{2-\beta_0^2}\,\frac{x_0^2}{r_0^2} \,.
\eq
Then
\bq
\begin{split}
\frac{d\alpha}{d\theta} &\simeq \frac{1}{\gamma_0} \bigl[\, 2\xi +(3-\beta_0^2)\xi^2 \,\bigr] 
\\
&= \frac{1}{\gamma_0} \biggl[\, 
\frac{2x_0}{r_0}\,\cos(\nu_x\phi) 
-\frac12\,\frac{6-5\beta_0^2}{2-\beta_0^2}\,\frac{x_0^2}{r_0^2} 
+(3-\beta_0^2) \frac{x_0^2}{r_0^2}\,\cos^2(\nu_x\phi) 
+\cdots
\,\biggr] 
\\
&= \frac{1}{\gamma_0} \biggl[\, 
\frac{2x_0}{r_0}\,\cos(\nu_x\phi) 
-\frac12\,\frac{6-5\beta_0^2}{2-\beta_0^2}\,\frac{x_0^2}{r_0^2} 
+\frac{3-\beta_0^2}{2} \frac{x_0^2}{r_0^2} +\cdots
\,\biggr] 
\\
&= \frac{1}{\gamma_0} \biggl[\, 
\frac{2x_0}{r_0}\,\cos(\nu_x\phi) 
+\frac12\,\frac{\beta_0^4}{2-\beta_0^2}\,\frac{x_0^2}{r_0^2} 
+\cdots
\,\biggr] 
\\
&= \frac{2}{\gamma_0}  \,\frac{x_0}{r_0}\,\cos(\nu_x\phi) 
+\frac12\,\frac{\gamma_0\beta_0^4}{\gamma_0^2+1}\,\frac{x_0^2}{r_0^2} 
+\cdots \,.
\end{split}
\eq
The secular term is
\bq
\label{eq:dadtsec}
\biggl\langle \frac{d\alpha}{d\theta} \biggr\rangle \simeq 
\frac12\,\frac{\gamma_0\beta_0^4}{\gamma_0^2+1}\,\frac{x_0^2}{r_0^2} \,.
\eq
This matches well with the tracking output. This is good.

\begin{figure}[!htb]
\centering
\includegraphics[width=0.95\textwidth]{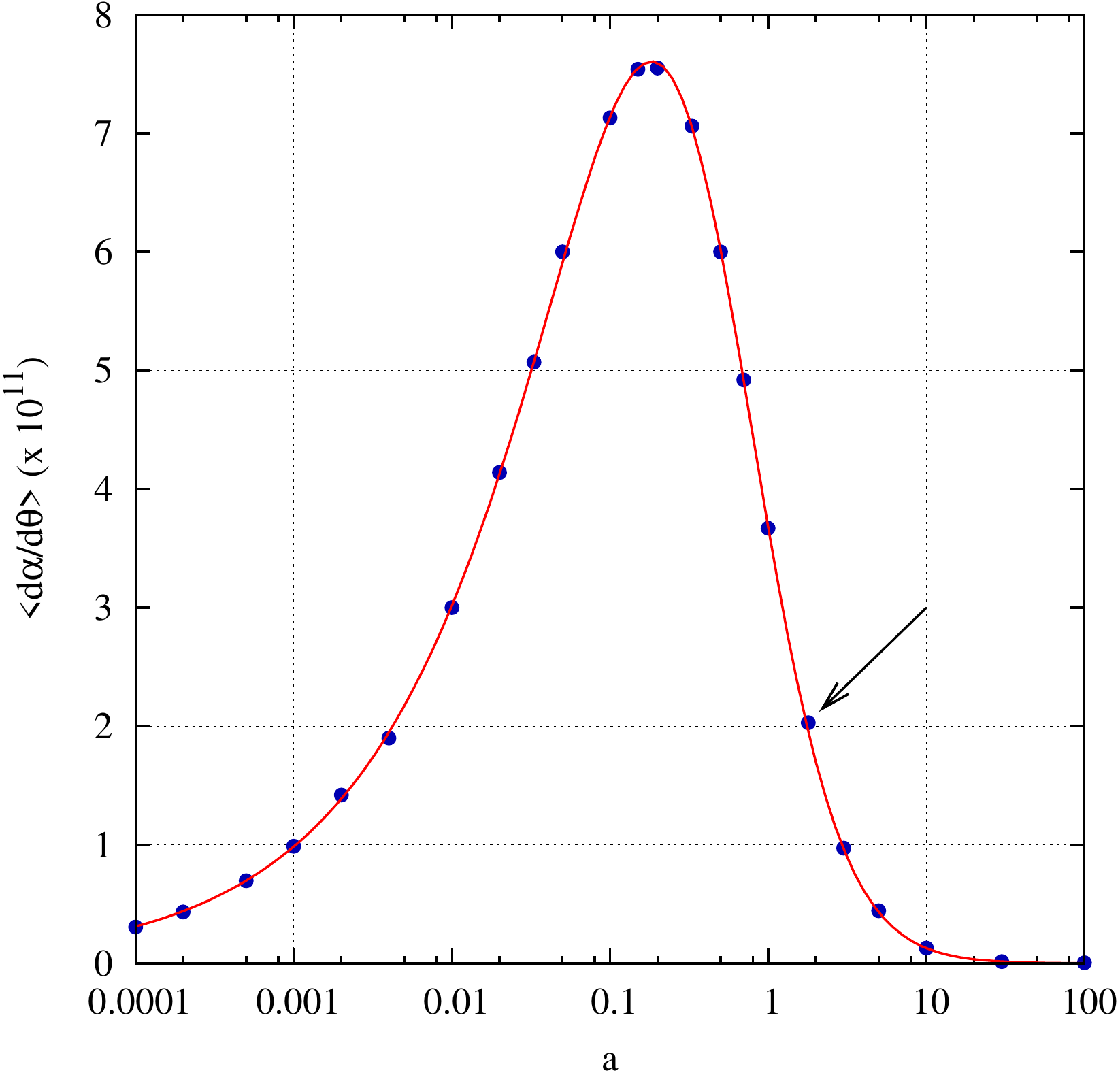}
\caption{\small
\label{fig:dadtavg}
Graph of $\langle d\alpha/d\theta\rangle$ v.~$a$ for a model homogenous weak focusing ring with a radial electric field.
The vertical scale has been multiplied by $10^{11}$ and the horizontal axis is logarithmic.
The circles represent the tracking data and the solid curve is the analytical formula.
The value for a proton is indicated with an arrow.
}
\end{figure}

\vskip 0.2in
Figure \ref{fig:dadtavg} displays a graph of 
$\langle d\alpha/d\theta\rangle$ vs.~$a$ for a 
model homogenous weak focusing ring with a radial electric field.
The ring radius was 40 m.
One particle was tracked for one million turns, 
with an initial value $x_0=1$ mm and $p_{x0}=0$.
The agreement between the tracking data and the analytical formula is excellent.

\section{Remarks}
Although eq.~\eqref{eq:dadtsec} is analytical,
it was really the tracking output which pointed to this expression.
Note that it was the {\em tracking program} which yielded the correct answer,
and the analytical theory then caught up, to fit the tracking results.
The initial theoretical analysis assumed that $\langle x_\beta\rangle=0$,
and led to the na{\"\i}ve expression eq.~\eqref{eq:dadtnaive}.
Indeed, I have assumed that $\langle x_\beta\rangle=0$
in all of my statistical averages for EDM calculations up to now.
I now know that this is not so.
I also know now that my tracking program is suffciently accurate and reliable to be taken seriously, 
if and when its output disagrees with analytical results.

\vskip 0.1in  
Basically, the phase space orbits in an EDM ring are not exactly ellipses centered on the origin.
The distortions of the phase space tori, though small, are significant enough to matter for EDM work.
Note that in general it is standard practice to assume that $\langle x_\beta\rangle=0$ when calculating statistical averages.
This is satisfactory in most applications.
Typically, the anharmonic terms (for example due to sextupoles) are small,
except in deliberate cases such as slow extraction.
However, for a logarithmic potential, such as in the EDM ring,
the anharmonic terms are sufficiently large that the value of 
$\langle x_\beta\rangle$ makes a significant contribution to
the secular rate of change of the helicity,
viz.~$\langle d\alpha/d\theta\rangle$.
This calls into question many assumptions which have been made about
statistical averages for EDM calculations.
Certainly, the statistical averages in my papers must be reexamined.

\vskip 0.1in  
All of the work reported in this note for was on-energy particles.
I did not treat an energy offset $H \ne H_0$.
I also did not treat models of electric fields for a different value of the field index, i.e.~non-logarithmic potentials.
I also did not treat vertical motion.
These are all issues for future work.

\vfill\pagebreak

\end{document}